\documentstyle[12pt,psfig]{article}

\begin{document}

\begin{flushright}
{VECC-NTH/98002}\\
\end{flushright}

\centerline{\large $\rho - \omega$ MIXING AT HIGH TEMPERATURE AND DENSITY}
\vskip 36pt
\centerline {\bf{Abhijit Bhattacharyya\footnote{Email : abhijit@veccal.ernet.in}}}
\vskip 1pt
\centerline{Variable Energy Cyclotron Centre}
\centerline{1/AF, Bidhannagar}
\centerline{Calcutta 700 064, INDIA}
\vskip 36pt
\vskip 15 pt

\begin {abstract}

The temperature and density dependence of the $\rho-\omega$ mixing amplitude 
has been studied from a purely hadronic model. The in-medium baryon masses 
and chemical potentials have been obtained at the Mean Field (MF) Level and 
these results have been used to calculate the mixing amplitude. It has been 
observed that the mixing amplitude changes substantially at high temperature 
and density.  
\end{abstract}

Charge Symmetry Violation (CSV) is a very well established and interesting 
sphere of research. Though this field of research is quite old it is still 
quite challenging as the finer corrections from the experiments are still 
coming. The cross section for the process $ e^+ e^- \rightarrow \pi^+ \pi^-$, 
in the $\rho - \omega$ resonance region, reveals an interference shoulder 
which results from the superposition of the narrow resonant $\omega$ and 
broad resonant $\rho$ amplitudes \cite{a}-\cite{c}. Thus a G-parity violating 
process $\omega \rightarrow \pi^+ \pi^-$ is observed. This observation 
motivates the study of $\rho - \omega$ mixing. Different authors have 
studied the $\rho - \omega$ mixing from different models and prescriptions 
\cite{d}-\cite{h1}. 
In this article the temperature and density dependence of the mixing 
amplitude will be studied. The $\rho - \omega$ mixing may be important 
in the context of hot and dense hadronic matter which can be observed in 
high energy heavy ion collisions. So it is interesting to see whether this 
amplitude is changed substantially at high temperature/density. 

After a brief discussion of the model the baryon masses and chemical 
potentials, in the hot and dense matter, will be calculated in the Mean 
Field Approximation (MFA) self consistently. This result will be used to 
calculate the temperature and density dependence of the mixing amplitude. 

The description of hadronic matter at high temperature should be based on a 
reliable relativistic model \cite{i}. The model should reproduce the basic 
features of the strong interaction {\it {i.e.}} short range repulsive and 
long range attractive forces. Here we will use non-linear Walecka model to 
calculate the mixing amplitude. 

The non-linear Walecka model Lagrangian with  nucleons, scalar mesons and 
vector ($\omega$ and $\rho$) mesons can be written as \cite{j}: 
\begin{eqnarray}
{\cal L} &=& \sum_B {\bar \psi}_B (\gamma^\mu p_\mu - m_B  
+ g_\sigma \sigma - g_\omega \gamma^\mu \omega_\mu) \psi_B  
+ {1 \over 2} (\partial^\mu \sigma \partial_\mu \sigma 
- m_\sigma^2 \sigma^2) \nonumber\\
&-& {1 \over 4} F_{\mu \nu} F^{\mu \nu} + {1 \over 2} m_\omega^2 \omega^2 
-{1 \over 3} b m_N (g_\sigma \sigma)^3 - {1 \over 4} c (g_\sigma \sigma)^4 
\nonumber\\
&-& {1 \over 4} G_{\mu \nu} G^{\mu \nu} + {1 \over 2} m_\rho^2 
\rho^2 
+ g_\rho {\bar \psi} \gamma^\mu {\tau.}\psi \rho_\mu 
+ f_\rho {\bar \psi} \sigma^{\mu \nu} { \tau.} \psi {{\partial_\mu} 
\over {2m}} \rho_\nu
\label{aa}
\end{eqnarray}

In the above expression $\psi_B, \sigma$, $\omega$ and $\rho$ 
are respectively the baryon, the $\sigma$-meson, the $\omega$-meson
and the $\rho$-meson fields; $m_B, m_\sigma$, $m_\omega$ and 
$m_\rho$ are the corresponding masses; $F_{\mu \nu} 
= \partial_\mu \omega_\nu - \partial_\nu \omega_\mu$ and 
$G_{\mu \nu} = \partial_\mu \rho_\nu 
- \partial_\nu \rho_\mu +[\rho_\mu, \rho_\nu]$. Here we have considered 
two baryons: neutron($n$) and proton($p$). Their masses are $m_n = 939.5731$ 
and $m_p = 938.2796$. In equation (\ref{aa}), $m_N = (m_n + m_p)/2$. 
The other parameter values used here are given by: $m_\rho = 770.0 MeV$, 
$m_\omega = 783.0 MeV$, $g^2_\rho/4\pi = 0.74$, $g^2_\omega/4\pi = 5.48$, 
$g_\sigma^2/4\pi = 5.35$ and 
$C_\rho = f_\rho/g_\rho = 6.1$ \cite{e,j1}. 

In ref. \cite{e} similar interactions of $\rho$ and $\omega$ have 
been used to calculate the momentum dependence of the $\rho - \omega$ 
mixing amplitude. The advantage of using this prescription, as pointed 
out in ref. \cite{e}, is that one can write down the contribution of 
the $\rho - \omega$ mixing to the $NN$ potential in a parameter free form: 
\begin{equation}
V^{\rho \omega}_{NN} (q) = - {{g_\rho g_\omega \langle \rho |H| 
\omega \rangle} 
\over {(q^2 - m_\rho^2)(q^2 - m_\omega^2)}}
\end{equation}
where the $\rho - \omega$ mixing amplitude is given by:
\begin{equation}
\langle \rho |H| \omega \rangle = g_\rho g_\omega \Pi (q)
\end{equation}

In the above equation $\Pi (q)$ is the transverse part of the polarisation 
function at finite temperature (the definetion of of $\Pi(q)$ is somewhat 
different compared to that in ref \cite{e}). 

The task ahead is now to calculate $\Pi (q)$ at finite temperature and
density keeping in 
mind that the $\rho$-meson has both vector and tensor coupling to the 
nucleons. So one has \cite{e}
\begin{equation}
\Pi^{\mu \nu} (q,T) = \Pi^{\mu \nu}_{vv} (q,T) + C_\rho \Pi^{\mu \nu}_{vt} 
(q,t)
\end{equation}
where
\begin{eqnarray}
i \Pi^{\mu \nu}_{vv} &=&  \int 
{{d^4k} \over {(2 \pi)^4}} Tr \left[\gamma^\mu G(k+q)  
\gamma^\nu { \tau_z} G(k) \right] \nonumber\\
i \Pi^{\mu \nu}_{vt} &=&  \int 
{{d^4k} \over {(2 \pi)^4}} Tr \left[\gamma^\mu G(k+q)  
{{i \sigma^{\nu \lambda} q_\lambda} \over {2m}} {\tau_z} G(k) \right] 
\end{eqnarray}
where $G$ is the finite temperature nucleon propagator.

The nucleon propogator can be splitted into its isoscalar and isovector 
components as: 
\begin{equation}
G(k) = {1 \over 2} G_p(k) (1 + \tau_z) + {1 \over 2} G_n(k) (1 - \tau_z)
\end{equation}

Once this splitting is introduced in the polarisation function and the 
isospin trace is carried out one can easily see that the $\rho - \omega$ 
mixing amplitude is proportional to the difference between the proton and 
the neutron loops (fig. 1) \cite{e} {\it {i.e.}} 
\begin{equation}
\Pi_{\mu \nu}(q) = \Pi^{(p)}_{\mu \nu} (q) - \Pi^{(n)}_{\mu \nu} (q)
\end{equation}

\begin{tabular}{cl}
\psfig{file=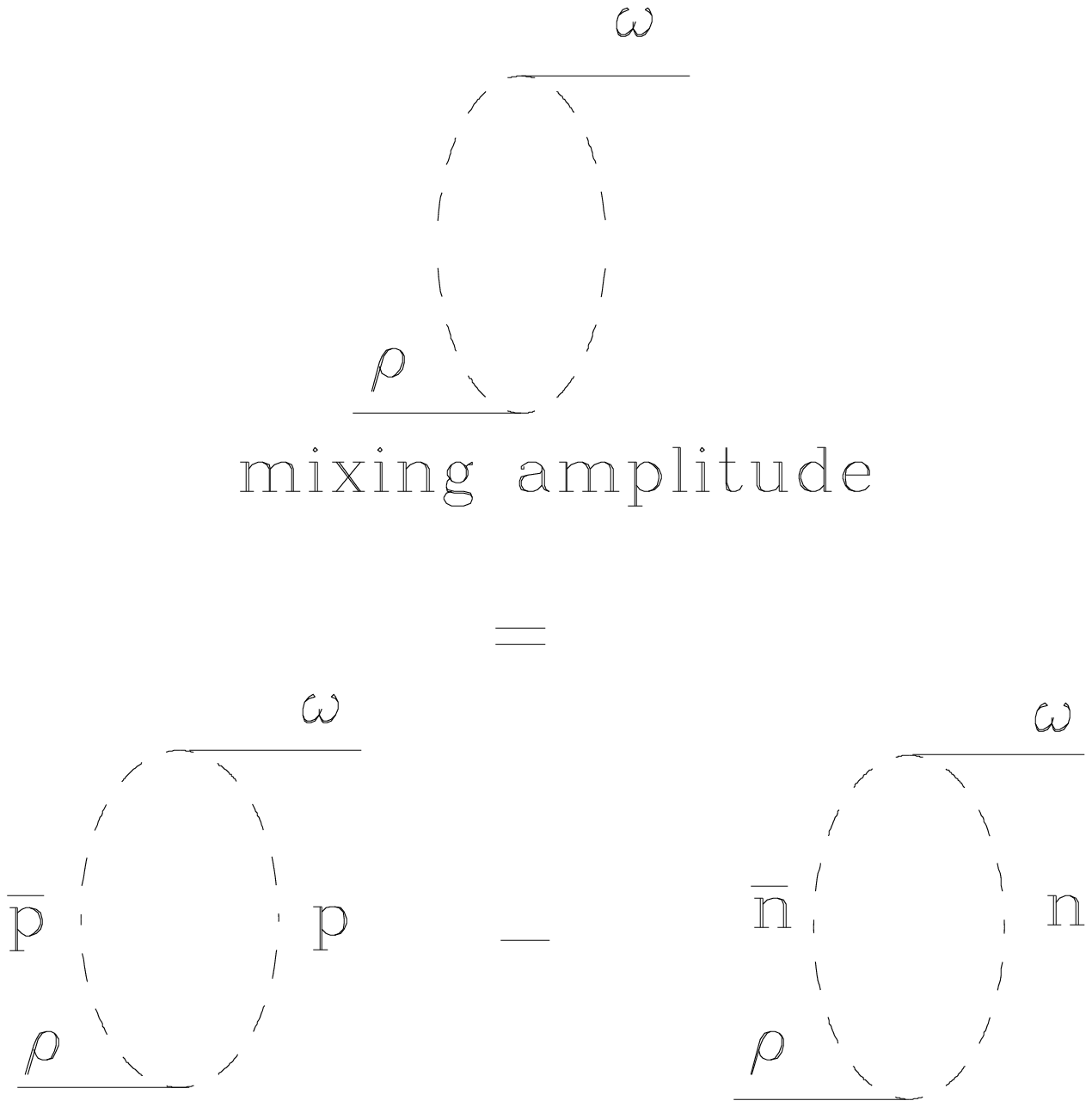,width=4in,height=3.5in}
\end{tabular}
\vskip 0.2in
\centerline{\bf Fig. 1 : Feynmann diagram for the  $\rho - \omega$
mixing amplitude.}
\vskip 0.2in

In the above set of equations the propagator for the baryons can be given by  
\begin{equation}
G_B(p) = (\gamma^\mu p_\mu + m_B^*) \left[{1 \over {(p^2 - 
m_B^{*2} + i \epsilon)}} + 2 \pi i \delta (p^2 - m_B^{*2}) sin^2 \phi_{p_0}
\right]
\end{equation}
where
\begin{eqnarray}
sin \phi_{p_0} &=& {{e^{-x/2} \theta (p_0)} \over {(1 + e^{-x})^{1 \over 2}
}}
- {{e^{x/2} \theta (-p_0)} \over {(1 + e^x)^{1 \over 2}}} \nonumber\\
x &=& (p_0 - \mu_B)/T
\end{eqnarray}
$B$ is either $p$ or $n$, $T$ is the temperature 
 and $\mu_B$ is the chemical potential.

Once we use this propagator the calculation of the polarisation function is 
really straight forward and we have \cite{e,i}
\begin{eqnarray}
\Pi^{(p)}_{vv} (q) &=& - {q^2 \over {2 \pi^2}} \left[{1 \over {6 \epsilon}} 
- {\gamma \over 6} - \int dx x(1-x) \left[{{m_p^2 - x(1-x)q^2} \over 
{\Lambda^2}}\right] \right] \nonumber\\
&-& 16 q^2 \int {{d^3k} \over {(2\pi)^3}} 
{{[f_p^+(E_p) + f_p^-(E_p)]} \over  {2E_p(k)}}
\left [ {{(E_p^2(k) - |k|^2 (1 - cos^2 \theta)/2)} \over {q^4 -4(k.q)^2}}\right] 
\nonumber\\
\Pi^{(p)}_{vt} (q) &=& - {q^2 \over {8 \pi^2}} \left [{1 \over {\epsilon}} 
- {\gamma} - \int dx \left[{{m_p^2 - x(1-x)q^2} \over {\Lambda^2}}\right] 
\right] \nonumber\\
&+& 8 q^4 \int {{d^3k} \over {(2\pi)^3}} 
{{[f_p^+(E_p) + f_p^-(E_p)]} \over  {2E_p(k)}}
\left[ {1 \over {q^4 -4(k.q)^2}}\right] 
\end{eqnarray}
and similar expressions for the neutron loops. In the above expression 
$f^\pm_B (E_B) = {1 \over {1 +exp((E_B \mp \mu_B)/T)}}$ is the thermal 
distribution function, $\Lambda$ is an arbitary renormalisation cutoff and 
$\gamma$ is the Euler-Mascheroni constant. Since we need to have the 
difference between the $p$ and $n$ loop contributions we get 
\begin{eqnarray}
\Pi_{vv} (q) &=& \Pi^{(p)}_{vv} - \Pi^{(n)}_{vv} \nonumber\\
&=& q^2 {1 \over {2 \pi^2}}  
\int dx x(1-x) \left[{{m_p^2 - x(1-x)q^2} \over {m_n^2 - x(1-x)q^2}}\right] 
\nonumber\\
&-& 16 q^2 \int {{d^3k} \over {(2\pi)^3}} 
\left[{{[f_p^+ + f_p^-]} \over  {2E_p}}
\times {{(E_p^2 - |k|^2 (1 - cos^2 \theta)/2)} \over {q^4 -4(k.q)^2}} \right .
\nonumber\\
&-& \left .{{[f_n^++f_n^-)]} \over  {2E_n}}
\times {{(E_n^2-|k|^2 (1 - cos^2 \theta)/2)} \over {q^4 -4(k.q)^2}}\right] 
\nonumber\\
\Pi_{vt} (q) &=& \Pi^{(p)}_{vt} - \Pi^{(n)}_{vt} \nonumber\\ 
&=& q^2 {1 \over {8 \pi^2}}  
\int dx \left[{{m_p^2 - x(1-x)q^2} \over {m_n^2 - x(1-x)q^2}}\right] \nonumber\\
&+& 8 q^4 \int {{d^3k} \over {(2\pi)^3}} 
\left[{{[f_p^+ + f_p^-]} \over  {2E_p}}
- {{[f_n^+ + f_n^-]} \over  {2E_n}}\right]
\times {1 \over {q^4 -4(k.q)^2}}
\label{bb}
\end{eqnarray}
where $E_B = \sqrt{k^2 + m^{*2}_B}$ is the effective energy of the baryon. 

From equation (\ref{bb}) it is clear that the mixing amplitude dependse
on the in-medium baryon masses and chemical potentials. In the MFA the 
effective baryon masses are : $m_B^* = m_B - g_\sigma \sigma_0$. 
The baryon number conservation gives $\rho = n_n + n_p$ where $\rho$ is
the total baryon number density and $n_n$ and $n_p$ are the number
densities of the neutrons and protons respectively. The asymetry in the
numbers of neutrons and protons is given by $x = (n_n - n_p)/(n_n +
n_p)$ where $x$ is the asymetry factor. In this paper, where a $Pb - Pb$
system has been considered, the asymmetry is $44/208$. From the above
relations one can find out the medium dependence of the
baryon masses and chemical potentials. 

We can now feed these in-medium baryon masses 
and chemical potentials in equation(\ref{bb}). After calculating the 
total polarisation function numerically we have plotted the absolute
value of the mixing amplitude 
$|\langle \rho |H| \omega \rangle|$ at $q^2 = m_\omega^2$ as a function of 
temperature, for different densities, in figure~2. 
The modulus of the mixing amplitude increases with temperature for 
$\rho = 0 fm^{-3}$. At $T = 0 MeV$ the amplitude is about $4300 MeV$ and
at $T = 200 MeV$ it is about $5000 MeV$. For $\rho = 0.15 fm^{-3}$, {\it
i.e.} at nuclear matter density, the amplitude first decreases from 
$30000 MeV$, at $T = 0 MeV$, to about $15000 MeV$, at $T = 180 MeV$.
Then it increaess slightly.  
At $\rho = 0.3 fm^{-3}$ the modulus of the mixing amplitude decreases
form $100000 MeV$ to $36000 MeV$ as the temperature is varried from $T =
0$ to $T = 200 MeV$. 
In fig.3 we have plotted the
density dependence of the mixing amplitude for different temperatures.
The amplitude increases sharply with density. But, with increase in temperature 
it decreases. The dominant contribution to the 
mixing amplitude, at low densities, comes from the 
vacuum part of the polarisation function which depends on 
temperature/density through the baryon 
masses only. At higher densities it comes from the part of the polarisation 
function which directly dependence on temperature/density through the 
distribution functions.

To conclude, we have calculated the temperature and density depenence of 
the $\rho - 
\omega$ mixing amplitude starting from a purely hadronic model. The mixing 
was assumed to be generated solely by $N {\bar N}$ loops and thus driven 
by the difference in the contribution from $n$ and $p$ loops. The mixing
amplitude is found to change substantially.
This can have a strong bearing on the QGP diagnostics.
The emission rate of dileptons ($l^+ l^-$) may get modified due to this
mixing. One can, however, use the temperature and density dependence of
baryon masses and chemical potentials from other models. This may change
the results quantitatively to some extent but, the qualitative behaviour
should not change. Studies in these directions are in progress. 

I thank Sibaji Raha for helpful discussions. The 
work has been supported partially by the Department of Atomic Energy 
(Government of India).
\newpage
\vskip 0.1in
\begin{tabular}{cl}
\psfig{file=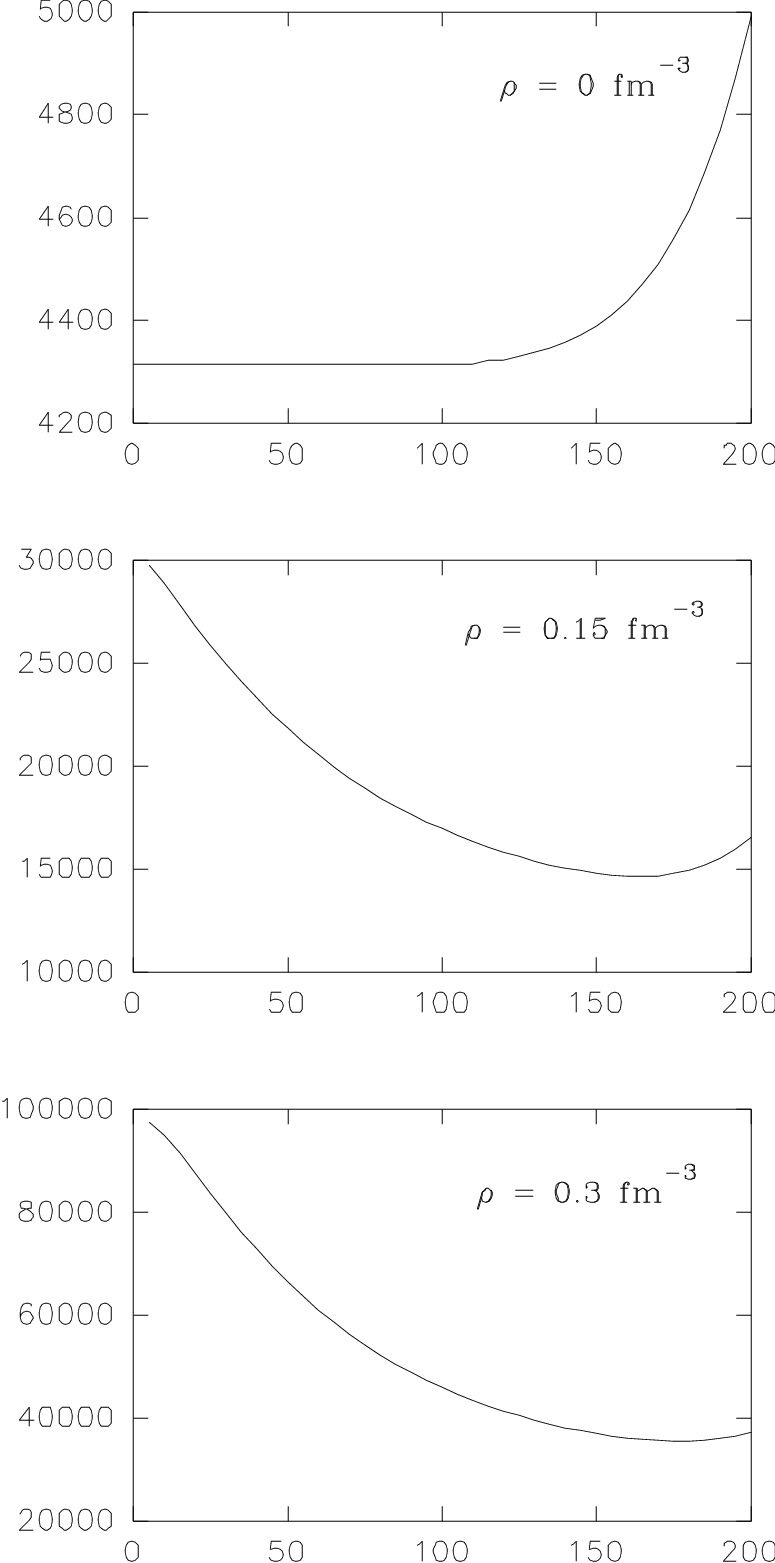,width=5in,height=7in}
\end{tabular}
\vskip 0.2in
\noindent
{\bf Fig. 2 : Temperature dependence of the modulus 
of the $\rho - \omega$ mixing amplitude ($|\langle \rho|H|\omega \rangle|$).}
\newpage
\vskip 0.1in
\begin{tabular}{cl}
\psfig{file=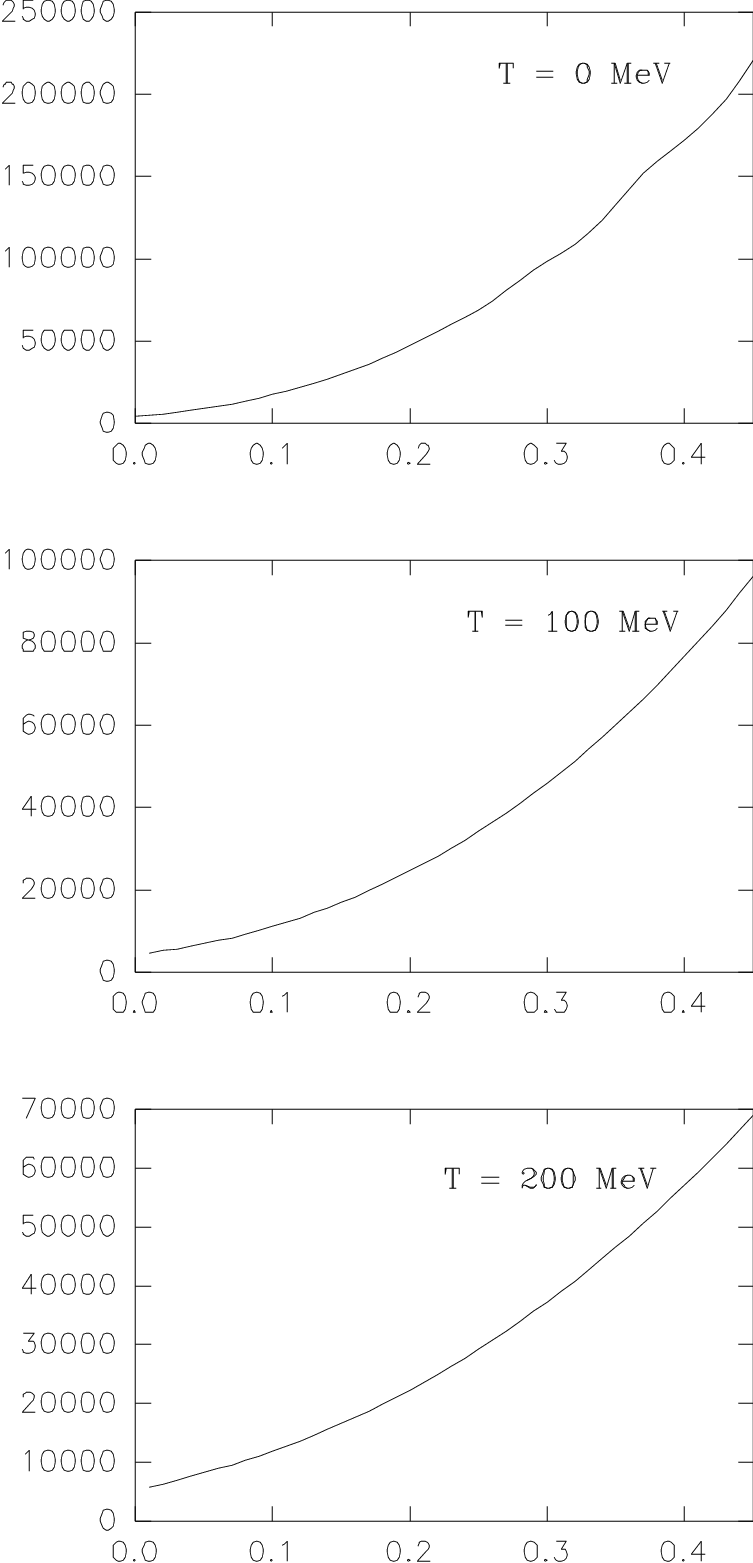,width=5in,height=7in}
\end{tabular}
\vskip 0.2in
\noindent
{\bf Fig. 3 : Density dependence of the modulus 
of the $\rho - \omega$ mixing amplitude ($|\langle \rho|H|\omega \rangle|$).}

\end{document}